\documentstyle[preprint,aps]{revtex}
\begin{document}
\tightenlines
\draft
\title{Analytical stripe phase solution for the Hubbard model.}
\author{S.I. Matveenko}
\address{Landau Institute for Theoretical Physics, Kosygina Str. 2, \\
117940, Moscow, Russia\\
and}
\author{S.I. Mukhin}
\address{Theoretical Physics Dept.,\\
Moscow Institute for Steel and Alloys, Leninskii pr. 4,\\
117936 Moscow, Russia}
\date{\today}
\maketitle
\begin{abstract}
The self-consistent solution for the spin-charge solitonic superstructure
in quasi-one-dimensional electron system is obtained in the framework of 
the Hubbard model as a function of a hole doping.
Effects of interchain interactions on the ground state are discussed.
Results are used for the interpretation of the observed stripe phases in doped
antiferromagnets.
\end{abstract}


It is well known \cite{gru} that at low enough temperatures
quasi-one-dimensional conductors may undergo a Peierls- or spin-Peierls
transition and develop accordingly either a long-range charge- or
spin-order. In the case of a discrete lattice model the commensuration
effects become important \cite{la}. Exact solutions of related Hartree-Fock
problems led to the picture of a solitonic lattice appearing away from
half-filling of the bare electron band \cite{br,machida}. The nature of a
soliton was determined then by a corresponding order parameter which was
either lattice deformation, or density of electronic spin. Nevertheless,
recently discovered stripe phases in doped antiferromagnets (cuprates and
nickelates) \cite{tranq} have attracted attention to the problem of
{\it coupled} spin and charge order pararmeters in the electron systems.
Numerical mean-field calculations \cite{zaan,schul} suggest a universality of
the spin-charge multi-mode coupling phenomenon in repulsive electronic systems
of different dimensionalities.
On the other hand, those calculations are bound to use small clusters which
often makes them inconclusive.

In this paper an exact analytical solution of the Hartree-Fock problem for a
one-dimensional electron system at and away from half-filling is presented.
This solution provides a unique possibility to investigate {\it analytically} the
structure of the periodic {\it spin-charge} solitonic superlattice.
It also demonstrates fundamental importance of the higher order
commensurability effects, which result in special stability points along the axis of
concentrations of the doped holes. Though there is no long-range order in the purely
one-dimensional system due to destructive influence of fluctuations, real cuprates
are three-dimensional, and therefore, the long-range order survives in the ground
state.
Hence, we believe,
that one-dimensional mean-field solutions contain universal features
of the  stripe phase, which are stabilized in higher dimensions. We use derived here
single-chain analytical solutions as building blocks for the stripe phase in quasi
two(three)-dimensional system of parallel chains. In this way
we have found that short-range (nearest neighbour) repulsion between doped
holes, in combination with effects of magnetic misfit energy between the
chains, naturally leads to formation of either ``half-filled'' or ``fully filled''
domain walls in the low- or high doping limits respectively.
In both cases these walls separate neighbouring antiphased antiferromagnetic
domains. The ``bi-stripe'' solution shows up as well.

The Hubbard Hamiltonian with the hopping integral $t$ and on-site repulsion
$U$ ($>0$) may be written in the form:

\begin{equation}
H=\displaystyle t\sum_{\langle i,j\rangle
\sigma}c^{\dagger}_{i,\sigma}c_{j,\sigma}+ U\displaystyle\sum_{i}\left(\frac{%
1}{4}\hat{n}^{2}_{i}-(\hat{S}^{z}_{i})^{2}\right)  \label{hubbard}
\end{equation}

\noindent Here an identity: $\hat{n}_{i\uparrow}\hat{n}_{i\downarrow}=\frac{1%
}{4}\hat{n}^{2}_{i}- (\hat{S}^{z}_{i})^{2}$ is used. Operators $\hat{n}%
\equiv\hat{n}_{\uparrow}+\hat{n}_{\downarrow}$ and $\hat{S}^{z}$ are
fermionic density and spin (z-component) operators respectively, and $\sigma$
is spin index. Hamiltonian (\ref{hubbard}) has convenient form for
Hartree-Fock decoupling in the presence of the two order parameters, i.e.
electron spin- and charge-densities , $m(x)$ and $\rho(x)$ respectively.
Single-particle eigenstates and eigenvalues of the Hamiltonian Eq.(\ref
{hubbard}) in the Hartree-Fock approximation can be determined from the
Bogoliubov-de Gennes equations derived in \cite{schul}:

\begin{equation}
\hat H \left( \begin{array}{c}
\Psi_{\sigma +} (x) \\
\Psi_{\sigma -}(x)
\end{array}
\right)
 = \left(
\begin{array}{cc}
-i\displaystyle\frac{d}{dx} +\alpha \rho (x) & -\alpha m(x) \\
-\alpha m(x)^* & i\displaystyle\frac{d}{dx}+\alpha \rho (x)
\end{array}
\right )\left(\begin{array}{c}
\Psi_{\sigma +} (x) \\
\Psi_{\sigma -}(x)
\end{array}\right) = E/(2t) \left( \begin{array}{c}
\Psi_{\sigma +} (x) \\
\Psi_{\sigma -}(x)
\end{array}
\right),
\label{h0}
\end{equation}

\noindent where $\alpha = U/4t $, the Plank constant is taken as unity, and
length is measured in the units of the lattice (chain) period $a$. In these
units momentum and wavevector are dimensionless, and velocity and energy
posses one and the same dimensionality.
The wave function is defined in terms of
the right- left-movers $\Psi_{\sigma\pm} $ as

\begin{equation}
\Psi_{\sigma }(x) =\Psi_{\sigma +} (x)\exp( i\pi x/2) + \sigma \Psi_{\sigma -} (x)
\exp(-i\pi x/2 ),
\label{psi}
\end{equation}
where $\sigma =\uparrow, \downarrow$, and the Fermi-momentum is $p_F= \pi \rho/2$.
In the case of half-filling the
average number of electrons per site equals $\rho = 1$, i.e. $p_F=\pi/2$.
The averages $m(x)$ and $\rho (x)$ are defined as:
\begin{equation}
 \langle \hat{n}(x) \rangle =   \rho (x), \qquad
 \langle \hat{S}^{z}(x)\rangle = m(x) \exp(i\pi x) + m^*(x) \exp(-i\pi x).
\label{def}
\end{equation}
In the case of half-filling: $\langle \hat{S}^{z}(i)\rangle = (-1)^i m(i)$, where
the order parameter $m(x)$ must be real.
The total energy is
equal to
\begin{equation}
\frac{W}{2t} = \sum_{E< \mu} \epsilon + \int dx \frac{\alpha}{2} (m^{*}(x)m(x) -
\rho^2 (x)),  \label{w}
\end{equation}
where $\epsilon = E/2t$, $\mu$ is a chemical potential.
Next, we introduce $\bar \rho $ and $\tilde \rho $ as
\[
\rho (x) = \bar \rho + \tilde \rho (x),\qquad \int\tilde\rho (x) dx = 0,
\]
and pass to a new basis $\Psi_1 $, $\Psi_2$, according to \cite{muk}, which in fact
is the Bloch-wave basis in the periodic charge-density wave potential
$U\tilde{\rho}(x)$
 (in what follows we drop spin index $\sigma$) :

\begin{equation}
\Psi_{\pm }(x) = \exp(\mp i\alpha \int^{x} \tilde \rho \, dx' ) \Psi_{1,2}(x),
\label{12}
\end{equation}

\noindent
Using this basis we can rewrite Eq.(\ref{h0}) in the form

\begin{equation}
\hat H \left( \begin{array}{c}
\Psi_1 (x) \\
\Psi_2(x)
\end{array}
\right)
= \left (
\begin{array}{cc}
-i\displaystyle\frac{d}{dx} & \Delta (x) \\
\Delta^*(x) & i\displaystyle\frac{d}{dx}
\end{array}
\right )\left(\begin{array}{c}
\Psi_1 (x) \\
\Psi_2(x)
\end{array}\right) = (E/2t - \alpha \bar\rho ) \left( \begin{array}{c}
\Psi_1 (x) \\
\Psi_2(x)
\end{array}
\right),
\label{h}
\end{equation}

\noindent
where

\begin{equation}
\Delta (x) = -\alpha m(x) \exp(2i\alpha \int\tilde \rho (x) dx ).
\label{delta}
\end{equation}

\noindent
It is easy to find from the finite band potential theory \cite{dubrovin} all
formal single-soliton solutions of the eigenvalue problem (\ref{h}):

\begin{equation}
\Delta (x) = \epsilon_0 -ik_0 \tanh k_0 x \; .
\label{dsol}
\end{equation}

\noindent
Here $\epsilon_0$ is the energy of the localized level, counted from the chemical
potential,

\begin{equation}
k_0 = \sqrt{\Delta_0^2 - \epsilon_0^2 }\; ,
\end{equation}

\noindent
and $2\Delta_0 $ is the gap in the energy spectrum.
Notice, that we consider $\Delta_0$ and $\epsilon_0$ (or equivalent pair of
variables) as {\it two} independent variational parameters. The reason is that,
according to Eqs. (\ref{h0}) and (\ref{delta}), there are {\it two
independent mean fields} ``hidden'' in $\Delta(x)$. Each of them should obey
a self-consistency equation. First, we derive equation for $m(x)$, using
Eq. (\ref{psi}):

\begin{equation}
m(x)+m^{*}(x)=\sum_{E< \mu}(\Psi_{+}(x)\Psi_{-}^{*}(x)+c.c.)
\label{spin}
\end{equation}

\noindent
which can also be rewritten in the $\Psi_{1,2}(x)$ representation 
Definition given in Eq. (\ref{12}) leads to a self-consistency equation for the
variable part of the charge mean-field
$\tilde\rho(x)\equiv \langle \hat{n}(x)\rangle-
\bar\rho$ :

\begin{equation}
\tilde \rho (x) = \sum_{E< \mu} (\Psi_1^+(x)\Psi_1(x) +
\Psi_2^+(x)\Psi_2(x))- \bar \rho   \label{rho}
\end{equation}

\noindent
Wave functions of the continuum spectrum are as follows:

\begin{equation}
\Psi_{1,2} =\frac{\pm \epsilon \mp \epsilon_0 +k +
ik_o \tanh k_0 x }{2\sqrt{L}\sqrt{\epsilon (\epsilon-\epsilon_0) -k_0 /L}}
e^{ikx},  \label{wv}
\end{equation}

\noindent
the upper (lower) sign on the r.h.s. corresponds to the index ``1''(``2'');
energy $\epsilon$ and wave vector $k$ are related by equation
\[
\epsilon^2 = \Delta_0^2 + k^2\; ,
\]
\noindent and $L$ is the length of the system (in units of the lattice spacing).

\noindent
Simultaneously, wave functions of the localized state with energy $\epsilon_0$ become:

\begin{equation}
\Psi_1(x) = \Psi_2(x) =\frac{\sqrt{k_0}}{2\cosh k_0 x }.  \label{lwf}
\end{equation}

\noindent
For correct summation over the energy levels in all the equations above, periodic
boundary conditions are imposed on the wave functions $\Psi_{\pm}$ of the continuum
spectrum:
\[
\Psi_{\pm}(x+L) =\Psi_{\pm}(x)\; .
\]

\noindent
Then, quantization condition follows:
\begin{equation}
kL + \arctan\frac{k_0}{k} =2\pi n,
\end{equation}
where n is an integer. Resulting self-consisting charge and spin components of the
single-soliton state are obtained:

\begin{eqnarray}
&&\displaystyle\tilde \rho (x)= \frac{k_0}{2\cosh^2 k_0 x }[\nu - 1 + \frac{2}{\pi}
\arcsin
\frac{\epsilon_0}{\Delta_0}],  \label{trho}\\
&&\displaystyle m(x) = - \frac{\epsilon_0 \pm ik_0 \tanh k_0 x }{\alpha }
\exp[\pm 2i\alpha (\nu - 1 +
\frac{2}{\pi} \arcsin \frac{\epsilon_0}{\Delta_0})(\frac{\tanh k_0x }{2} -
\frac{x}{%
L})]\exp(i\chi) ,  \label{m}
\end{eqnarray}

\noindent
where $\nu $ is the filling factor of the localized level ($\nu = 0, 1, 2 $); and
$\chi $ is an arbitrary phase.
As is evident from Eq. (\ref{def}), at half filling, i.e. in the commensurate case
$\rho = 1$, the order parameter $m(x)$ must be real.  In order to fulfil this
condition in Eq. (\ref{m}) (to the lowest order in $\alpha$ for the case $\nu =0, 2$,
and precisely, for $\nu = 1$) one chooses: $\epsilon_0=0$. Eqs. (\ref{trho}),
(\ref{m}) describe structures of the topological spin-charge solitons (kinks), which are
either {\it spinless} with charge $\pm e$ (single electron charge) at $\nu = 0,2$, or
{\it chargless} with spin $1/2$ in the case $\nu=1$. In all the cases there are two
antiphased antiferromagnetic domains in the system.  The  $\nu=1$ soliton is the
stationary excited state of the undoped system. The same is true in the cases $\nu=0$
($\nu =2$), but with one electron removed(added) from(to) the system (i.e. still
zero doping in the thermodynamic limit). Important is
that when doping with holes or electrons becomes finite, already the {\it ground state}
of the system  possesses periodic ``chain'' of the alternating spinless solitons
and antisolitons of the kind $\nu =0$ or $\nu =2$ (for hole- or electron doping
respectively), which are described above. This
solitonic superlattice, as we
assume, is a one-dimensional analogue of the stripe phase observed in lightly doped
cuprates and nickelates \cite{tranq}. Existence of the superlattice is
related to appearance of the central band around $\epsilon=0$, filled with either
spinless ``holes'' ( $\nu =0$) or spinless ``electrons'' ( $\nu =2$).

In order to
find a structure of the ground state at finite doping we calculate the total
energy Eq. (\ref{w}) of the electron system using solutions (\ref{trho}) and (\ref{m}).
We obtain for the potential energy:
\begin{equation}
\frac{W_{pot}}{2t} =\int dx \frac{\alpha}{2} (m^{*}(x)m(x) -{ \rho}^2 (x))= \frac{%
L\Delta_0^2}{2\alpha } -\frac{k_0}{\alpha} -\frac{\alpha }{6}(\nu -1 + \frac{%
2}{\pi} \arcsin\frac{\epsilon_0}{\Delta_0})^2k_0-\frac{\alpha}{2}L\bar\rho^2.
\label{wpot}
\end{equation}
The other part of the total energy reads:
\begin{equation}
\frac{W_{el}}{2t} =\sum_{E< \mu} \epsilon_i =-\displaystyle\frac{L}{\pi}(p_F\epsilon_F
 +\Delta_0^2 \ln \frac{\epsilon_F+p_F}{\Delta_0} ) +w,  \label{wel}
\end{equation}
where $  \epsilon^2_F = p^2_F + \Delta^2_0$, and
the term $w$ is of the order $L^0$.

Minimization of the total energy with respect to $\Delta_0 $ in the order $%
\propto L$ gives the usual result \cite{br} :
\begin{equation}
\Delta_0 = 2\epsilon_F \exp(-\frac{1}{\lambda}), \quad \lambda =\frac{2\alpha}{\pi}
\;.  \label{Delta}
\end{equation}

Since our approximation makes sense when $\Delta_0 \ll \epsilon_F$, we
conclude
that parameter $\alpha = U/4t$ should be much less than 1 (weak interaction
limit). In this limit we see that part in the potential energy $\int\alpha
\tilde\rho^2 /2$ is of the order $\alpha $ and is much less than other
terms, which are of the order $\alpha^{-1}$ or $\alpha^0 $.

The total energy of the kink
 state Eq. (\ref{w}) can be expressed similar to the work \cite{br}:
\begin{equation}
\frac{W -W_0}{2t} = \Delta_0 [(\nu -\frac{2\theta}{\pi})\cos \theta + \frac{2%
}{\pi} \sin \theta ] -\frac{\alpha}{6} \Delta_0(\nu -\frac{2\theta}{\pi}%
)^2\sin \theta ,  \label{wex}
\end{equation}
where $W_0$ is the energy of the state without a soliton,
$\cos \theta
=\epsilon_0/\Delta_0$, $\sin \theta = k_0/\Delta_0 $.
As we have already seen, $\epsilon_0 = 0$ for the half-filled case, thus leading to
$\theta = \pi /2$.

Minimization of (\ref{wex}) with respect to $\theta$ in the case $\rho \neq 1$
(it is assumed that $\vert\rho -1\vert \gg \Delta_0 /v_F $) leads to the equation:
\begin{equation}
\frac{2}{\pi} (\theta - \frac{\pi}{2}\nu ) [ (1-\frac{2\alpha }{3\pi}) \sin
\theta - \frac{\alpha}{3\pi}(\theta -\frac{\pi}{2}\nu )\cos \theta ]=0.
\end{equation}
Solutions are the same as for the Peierls model and are independent of $\alpha$ when
$\alpha \ll 1$ . The only nontrivial solution is a solitonic excited state with:
\begin{equation}
\nu = 1, \qquad \epsilon_0 \approx 0, \qquad q\approx 0, \qquad W-W_0 =2\Delta_0/\pi ;
\label{qq}
\end{equation}
where $q$ is the charge of a soliton. This chargeless solitonic excitation corresponds 
to a gradual phase-shift by $\pi/2$ of the argument of the sine function in the 
periodic ground state solution given below in Eq. (\ref{m2}).
Thus, we see that the
structure of the ground state of
our model should be similar to the one of  the Peierls model \cite{br}. Hence,
we can conclude that  for the finite hole density $n_h = \vert \rho - 1\vert$
charged  ($\nu = 0,\ 2$ ) solitons form a periodic structure with the spin
period $l= 2/\vert \rho - 1\vert $ equal twice the charge period.

In the lowest order expansion in $\alpha \ll 1$
this structure is described in terms of elliptic functions:
\begin{eqnarray}
&& \tilde \rho (x)= \frac{2K(r^{\prime})\Delta_0 r}{\pi}[
sn^2 (\Delta_0  x/r , r) -\frac{1}{r^2}(1-\frac{E(r)}{K(r)})],  \label{trho1}\\
&&m(x) = \Delta_0 \sqrt{k} sn(\Delta_0  x/  \sqrt{k} , k), \label{m1}
\end{eqnarray}
where $ sn (\Delta_0  x/  \sqrt{k} , k )$ is the Jakobi elliptic function
with the parameter $0<k<1$ defined by $2K(k)\sqrt{k}/\Delta_0 =
1/\vert \rho - 1\vert $, where $K(r)$ and $E(r)$ are complete elliptic integrals
 of the first and second kind respectively, and $r = 2\sqrt{k}/(k+1)$,
$r^{\prime}=\sqrt{1-r^2}$. Parameter $k$ varies from $k=1$ at $\rho =1$ where
$ sn(\Delta_0 x) = \tanh (\Delta_0 x)$, to $k \ll 1$ where
$sn (\Delta_0 x/\sqrt{k}, k) \sim \sin(\pi \vert\rho -1 \vert x)$. Simultaneously,
$E(0)/K(0)=1$, and $E(1)/K(1)=0$.

In the limiting case of ``overdoping'': $\vert \rho -1\vert \gg \Delta_0$, in which
case $k\ll 1$ and $K(k)\approx \pi/2$, one has:

\begin{equation}
 m(x) \approx    \frac{\Delta_0^2 }{\pi \vert \rho -1\vert}
 \sin (\pi \vert\rho - 1 \vert x),
\label{m2}
\end{equation}

\begin{equation}
\tilde{\rho}(x) \approx \frac{\Delta_0^4}{\pi^2\vert\rho -1\vert^3}
\cos (2\pi\vert\rho -1\vert x),
\label{trho2}
\end{equation}
in accord with the results of \cite{muk} obtained in the limit of main harmonics 
coupling.

It is known that in any exactly integrable model there is no commensurability
effects at the commensurate points \cite{dz} :

\begin{equation}
\vert \rho_0 - 1\vert = m/n,  \label{cs}
\end{equation}
where $m$, $n$ are relatively prime integers. That is the energy and other
system parameters continuously depend on the filling factor $\rho $. But in
our case the term $-\alpha \rho^2/2$ in the potential energy (\ref{w})
violates exact integrability and leads to a pinning of the spin density wave
$m(x)$. As a result at any rational point we obtain a decrease in the total
energy of the order \cite{dz}
\begin{equation}
\delta w \propto -\alpha e^{-n\,\, const }.
\end{equation}

To summarise we consider briefly the two-dimensional (2D) case using our 1D results by
adding weak interchain interactions. In the lowest order approximation in the
interchain
hopping integral $t_{\perp}$ the interaction energy is

\begin{equation}
\delta W=-J\sum_{<i,j>}\int_{0}^{L}dx(\cos (\varphi _{i}-\varphi
_{j})-1)+W_{Coulomb},  \label{W}
\end{equation}
where $J\sim t_{\perp }^{2}/\Delta _{0}$, $\varphi _{i}$ is the phase of a
spin-density $m(x)$ on the $i$-th chain, $W_{Coulomb}$ is the Coulomb interaction
energy between the charged kinks (solitons). We suppose
for simplicity that the Coulomb interaction decreases rapidly with the distance
and take into account only charged kinks on the neighbouring
sites of the neighbouring chains, $W_{Coulomb}=N_{p}q^{2}/\epsilon a'$, where
 $N_{p}$ is the number of pairs of the charged kinks, $\epsilon $ is a dielectric
constant, and $a'$ is an interchain spacing. In the half-filled system, i.e. in
the absence of the charged kinks, the minimum
of (\ref{W}) is
achieved when $\varphi _{i}=\varphi _{j}$ for all $i$, $j$, and we have $\delta W=0$.
There are two possible ways to create a periodic solitonic structure (solitonic
superlattice) in the doped 2D system. In the first case the charged kinks would reside
on every chain, while in the second case only every even (odd) chain would be doped.
Compare the energies of these two different configurations.

In the first case we have $\varphi_i = \varphi_j$ and
\begin{equation}
\delta W_1 = W_{Coulomb} = N_h q^2/2\epsilon a , \label{w1}
\end{equation}
where $N_h$ is the number of kinks (solitons) and the charge $q$ could be deduced from
Eqs. (\ref{trho}) and (\ref{trho1}). This is a 2D stripe pattern with filled
"domain walls", i.e. having
one spinless charged kink per period $a'$ perpendicular to the chains direction.
In the second case the Coulomb repulsion energy is negligible, but there is
an increment in the total energy due to the first term in (\ref{W}), i.e. due to a
magnetic order misfit: $\varphi_i \neq \varphi_j$. The minimal energy
configuration could be achieved now in two ways. One possible pattern corresponds to
half-filled "bi-stripe" pattern,  seen experimentally in some doped manganites
\cite{tranq}.
Namely, the kink - anti-kink pairs of the smallest possible size $\xi$ would form on
each even(odd) chain in order
to keep the phase-shift of $m(x)$ with respect to the odd(even) chains equal to
$2\pi$. The odd(even) chains remain antiferromagnetically ordered.
Then the energy is:
\begin{equation}
\delta W_2 \sim N_h J\xi ,\label{w2}
\end{equation}
where $\xi$ is of the order of the kink width. $\xi$ monotonically increases
from its value $v_F /\Delta_0 \sim e^{1/\lambda }$ at $\rho =1$, to $v_F
/\Delta \sim e^{2/\lambda }\tan \pi \vert \rho -1\vert/2 $ in the limit
$\vert \rho -1\vert \gg e^{-1/\lambda }$ \cite{ma}.
In the case $\delta W_{1}>\delta W_{2}$ we have that the alternating kink
structure is energetically preferable. Notice, that function $\delta W_{1}$
monotonically decreases, while function $\delta W_{2}$ monotonically
increases as the function of $|\rho -1|$, therefore the sign in the above
inequality may change at some sufficient doping concentration of the holes.

An alternative, half-filled single-stripe pattern may arise when
 $J\xi\gg 2\Delta/\pi$ , i.e. when $t_{\perp}$ is not too small.
In this case the minimum of the energy (\ref{W}) is achieved at $%
\varphi _{i}=\varphi _{j}$. For this to be true for any couple of the chains, while
$W_{Coulomb}\approx 0$ being also fulfilled, the doped holes should again reside, say,
on every even chain in the form of spinless charged solitons $\nu =0$ . But
simultaneously, an equal number of the chargeless solitons
$\nu =1$ (with spin $\pm 1/2$) must
be formed at all the odd chains in order to maintain
the condition $\varphi _{i}=\varphi _{j}$.
This
configuration will be stable if:
\begin{equation}
W_{s}N_{h}<\frac{q^{2}}{2\epsilon a}N_{h},  \label{x}
\end{equation}
where $W_{s}=2\Delta/\pi$ is creation energy of the chargeless kink.
Notice that since the charge $q$ monotonically decreases from 1 to 0 as function of
doping $N_{h}=L\vert 1-\rho\vert$ (see Eqs. (\ref{trho}),(\ref{trho2})), therefore the
inequality (\ref{x}) will be not satisfied at high doping densities, and half-filled to
filled stripe transition would be expected in qualitative accord with experiment
\cite{tranq}.

Instructive discussions with Jan Zaanen and S.A. Brazovski are highly
appreciated by the authors.

\end{document}